\documentclass{Interspeech}
\usepackage{fix-cm}
\usepackage{threeparttable}
\usepackage{graphicx}
\usepackage{subfigure}
\usepackage{multirow}
\usepackage{setspace}
\usepackage{booktabs}
\usepackage{cite}
\usepackage{setspace}
\usepackage{makecell}
\usepackage{ulem}
\usepackage{xcolor}
\usepackage[section]{placeins}
\usepackage{pifont}
\usepackage{hyperref}
\def\H{{\mathsf H}}

\def\CC{{\mathbb C}}

\usepackage{hyperref}

\interspeechcameraready

\title{ARiSE: Auto-Regressive Multi-Channel Speech Enhancement}

\author[affiliation={1,2}]{Pengjie}{Shen}
\author[affiliation={1}]{Xueliang}{Zhang}
\author[affiliation={2}]{Zhong-Qiu}{Wang}

\affiliation{School of Computer Science}{Inner Mongolia University}{China}
\affiliation{Department of Computer Science and Engineering}{Southern University of Science and Technology}{China}
\email{shenpengjie@mail.imu.edu.cn, cszxl@imu.edu.cn, wang.zhongqiu41@gmail.com}

\keywords{Auto-regressive multi-channel speech enhancement, beamforming, neural speech enhancement}

\begin{document}

\maketitle

\begingroup
\renewcommand\thefootnote{}\footnote{This work was done while Pengjie Shen was a visiting student at SUSTech. \textit{Corresponding author: Zhong-Qiu Wang}.}
\addtocounter{footnote}{0}
\endgroup

\begin{abstract}
We propose ARiSE, an \underline{a}uto-\underline{r}egress\underline{i}ve algorithm for multi-channel \underline{s}peech \underline{e}nhancement.
ARiSE improves existing deep neural network (DNN) based frame-online multi-channel speech enhancement models by introducing auto-regressive connections, where the estimated target speech at previous frames is leveraged as extra input features to help the DNN estimate the target speech at the current frame.
The extra input features can be derived from (a) the estimated target speech in previous frames; and (b) a beamformed mixture with the beamformer computed based on the previous estimated target speech.
On the other hand, naively training the DNN in an auto-regressive manner is very slow.
To deal with this, we propose a parallel training mechanism to speed up the training.
Evaluation results in noisy-reverberant conditions show the effectiveness and potential of the proposed algorithms.
\end{abstract}

\section{Introduction}

Multi-channel speech enhancement is an important task in speech signal processing.
Compared with single-channel cases, it usually produces better enhancement by leveraging spatial information afforded by multiple microphones \cite{Gannot2017ACP, Haeb-Umbach2019, Haeb-Umbach2025, Wang2021MCCSM, Tan2022NSF, Araki2025, Wang2022GridNetJournal}.
It has many applications in speech communication such as audio-video conferencing, human-machine voice interaction, and hearing aids \cite{Gannot2017ACP, Haeb-Umbach2019, Haeb-Umbach2025, Wang2021MCCSM, Tan2022NSF, Araki2025, 665, 5675860}.
Despite its importance, the field still faces major challenges, including effectively exploiting spatial information and robustly dealing with room reverberation \cite{reverberation}, speech interferences \cite{noise,interfer} and environmental noises \cite{noise,noise-and-interfer}.
These challenges make multi-channel speech enhancement a prominent focus of ongoing research in speech signal processing.


Recent advances in deep learning have revolutionized beamforming techniques for multi-channel speech enhancement \cite{WDLreview, Haeb-Umbach2019, Haeb-Umbach2025}. By leveraging DNNs to model complex signal patterns and optimize spatial feature extraction, modern beamformers can dynamically adapt to challenging acoustic environments with non-stationary noises, interferences, and reverberation \cite{deeplearn-based-beamforming,9053092}. This integration enables more accurate time-frequency (T-F) masking, resulting in remarkable improvements in speech quality and intelligibility compared to conventional methods. The core innovation lies in T-F masking based beamforming \cite{mask-beam,mask-beam2,att-beam,att-beam-self}, where DNN-generated T-F masks are leveraged to estimate spatial covariance matrices (SCMs) and these SCMs can be used to drive established beamformers such as minimum variance distortionless response (MVDR) \cite{665, Souden2010} and multichannel Wiener filter (MWF) \cite{SDW-MWF}. By precisely characterizing spatial relationships through the T-F masks estimated by DNNs, the approach can effectively separate target speech from interfering signals while maintaining acoustic fidelity and introducing little distortion to target signals.

Integrating auto-regressive (AR) models into existing frame-online speech enhancement models offers a promising avenue for capturing temporal dependencies within speech signals \cite{Recurrent-Deep-Stacking, DBLP:conf/interspeech/AndreevBSSA23, autoregressive-sep, Online-Speaker-aware-sep, pan24_interspeech}.
AR models enable more effective utilization of contextual information by leveraging the outputs generated in previous time steps to improve the prediction at the current time step.
This capability is particularly helpful in speech processing, where nearby frames in speech typically exhibit strong correlation.
Existing research indicates that using AR mechanisms can yield better 
performance for single-channel speech enhancement \cite{Recurrent-Deep-Stacking, DBLP:conf/interspeech/AndreevBSSA23} and single-channel speaker separation \cite{autoregressive-sep, pan24_interspeech}.
Although AR models may be difficult and slow to train due to the auto-regressive connections (i.e., the unfolded network during optimization would be extremely deep and hence hard to optimize), there are carefully-designed training algorithms \cite{Recurrent-Deep-Stacking, pan24_interspeech} capable of mitigating this issue.

In this context, we propose ARiSE, an auto-regressive algorithm for multi-channel speech enhancement.
Different from existing auto-regressive approaches \cite{Recurrent-Deep-Stacking,DBLP:conf/interspeech/AndreevBSSA23,autoregressive-sep,Online-Speaker-aware-sep, pan24_interspeech}, which are only designed for single-channel speech enhancement and speaker separation, we extend auto-regressive modeling to multi-channel scenarios.
In single-channel cases, only the DNN-estimated target speech in previous frames is used as the auto-regressive connections \cite{Recurrent-Deep-Stacking,DBLP:conf/interspeech/AndreevBSSA23,autoregressive-sep,Online-Speaker-aware-sep, pan24_interspeech}, while in multi-channel cases beamforming results (computed based on the estimated target speech in previous frames) can be additionally leveraged as auto-regressive connections.
In our experiments, we observe that, in multi-channel cases, beamforming results are much more effective auto-regressive connections than the DNN-estimated target speech in previous frames.

The contributions of this paper are summarized as follows:
\begin{itemize}[leftmargin=*,noitemsep,topsep=0pt]
\item We integrate T-F masking based beamforming in an auto-regressive manner for multi-channel speech enhancement, where a beamformer, computed based on the estimated target speech in previous frames, is leveraged to improve the estimation of the target speech at the current frame.
\item We design an effective way to speed up the training of the proposed auto-regressive enhancement models, following the RDS \cite{Recurrent-Deep-Stacking} and PARIS \cite{pan24_interspeech}  algorithms.
\end{itemize}
Evaluation results on multi-channel speech enhancement in noisy-reverberant conditions show the effectiveness of the proposed algorithms.

\section{Proposed Algorithms}


In a noisy-reverberant room with a single target speaker and a microphone array with $M$ microphones, the recorded mixture can be formulated, in the short-time Fourier transform (STFT) domain, as follows:
\begin{equation}
    \mathbf{Y}(t,f) = \mathbf{X}(t,f) + \mathbf{N}(t,f) \in \CC^M,
\end{equation}
where $\mathbf{Y}(t,f), \mathbf{X}(t,f)$ and $\mathbf{N}(t,f)$ respectively denote the captured noisy-reverberant mixture, target direct-path speech of the speaker, and non-target signals captured at time frame $t$ and frequency $f$.
Without loss of generality, microphone $q\in\{1,\dots,M\}$ is designated as the reference microphone.
We aim to estimate the target direct-path signal captured at the reference microphone (i.e., $X_q$) based on the multi-channel input mixture $\mathbf{Y}$ in a frame-online way.
Note that when the indexes $t$, $f$ or $q$ are dropped, we refer to the corresponding spectrogram.

This section first proposes
ARiSE and then investigates an effective training mechanism for it.

\begin{figure}[t]
    \centering
    \includegraphics[width=1.0\linewidth]{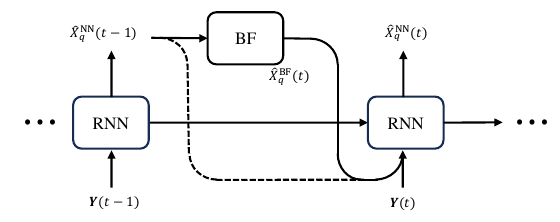}
    \caption{Illustration of proposed ARiSE algorithm. The connection in dashed line is optional.}
    \label{fig:enter-label}
\end{figure}

\subsection{ARiSE}
ARiSE is illustrated in Figure \ref{fig:enter-label}.
It operates in a frame-online manner, processing incoming mixtures frame by frame to estimate target speech.
The enhancement model can be any frame-online enhancement models such as a recurrent neural network (RNN).
At each time frame $t$, besides taking in the mixture at the current frame (i.e., $Y(t)$) as input, the model also uses as input (a) the estimated target speech produced by the model at the previous frame; and (b) beamformed mixture computed by leveraging the estimated speech in previous frames.
This mechanism, which leverages estimated target speech in previous frames to improve the estimation of the target speech at the current frame, forms a novel auto-regressive approach for multi-channel speech enhancement.

The beamformer is computed in a frame-online manner and by leveraging T-F masking \cite{mask-beamforming1,mask-beam2,mask-beamforming3}, where the T-F masks can be computed based on the estimated target speech in previous frames.
In detail, at time frame $t$, given estimated target speech in previous frames, we compute an MVDR beamformer \cite{Souden2010, Haeb-Umbach2019, Haeb-Umbach2025} at time frame $t-1$ as follows:
\begin{equation}
    \hat{\mathbf{w}}_q(t-1,f)=\frac{\hat{\boldsymbol{\Phi}}_{\text{N}}(t-1,f)^{-1} \hat{\boldsymbol{\Phi}}_{\text{X}}(t-1,f)}{\operatorname{Tr}\Big(\hat{\boldsymbol{\Phi}}_{\text{N}}(t-1,f)^{-1} \hat{\boldsymbol{\Phi}}_{\text{X}}(t-1,f)\Big)} \mathbf{u}_q, \label{MVDR_equation}
\end{equation}
where $\operatorname{Tr}(\cdot)$ is the trace operation and $\mathbf{u}_q\in\CC^M$ a one-hot vector with the $q^{\text{th}}$ element being one.
$\hat{\boldsymbol{\Phi}}_{\text{X}}(t-1,f)$ and $\hat{\boldsymbol{\Phi}}_{\text{N}}(t-1,f)\in\CC^{M\times M}$ are respectively the SCMs of the target and non-target signals.
They are computed as follows:
\begin{align}
\hat{\boldsymbol{\Phi}}_{\text{X}}(t-1,f) &= \sum_{t'=1}^{t-1} \hat{\mathbf{X}}^{\text{NN}}(t',f) \hat{\mathbf{X}}^{\text{NN}}(t',f)^{\H}, \\
\hat{\boldsymbol{\Phi}}_{\text{N}}(t-1,f) &= \sum_{t'=1}^{t-1} \hat{\mathbf{N}}(t',f) \hat{\mathbf{N}}(t',f)^{\H}, 
\end{align}
where $(\cdot)^{\H}$ denotes Hermitian transpose, $\hat{\mathbf{N}}=\mathbf{Y}-\hat{\mathbf{X}}^{\text{NN}}$, and $\hat{X}_m^{\text{NN}}(t',f)=Y_m(t',f)\times \hat{Z}_q(t',f)$ with $m\in\{1,\dots,M\}$ indexing the $M$ microphones and $\hat{Z}_q$ denoting the DNN-estimated complex-valued T-F mask at the reference microphone $q$.
The beamforming (BF) result at time frame $t$ and frequency $f$ can be computed in two ways:
\begin{align}
    \text{{BF option 1}}: \hat{X}_q^{\text{BF}}(t,f) &= \hat{\mathbf{w}}_q(t-1, f)^{\H} \mathbf{Y}(t-1, f) \label{option_1} \\
    \text{{BF option 2}}: \hat{X}_q^{\text{BF}}(t,f) &= \hat{\mathbf{w}}_q(t-1, f)^{\H} \mathbf{Y}(t, f) \label{option_2}
\end{align}
Notice the subtle differences between the two options.
In Equation \ref{option_1}, the beamformer computed at the previous frame is applied to the mixture at the previous frame, and the result is considered as the beamforming result for the current frame.
Differently, in Equation \ref{option_2}, the beamformer computed at the previous frame is applied to the mixture at the current frame.
The rationale of preferring Equation \ref{option_2} over Equation \ref{option_1} is that (a) the beamformer weights would be typically stable in consecutive frames, as long as the target speaker does not move drastically; and (b) the beamforming result computed in Equation \ref{option_2}, which is usually time-aligned to the current frame, would be more informative for enhancing the target speech at the current frame.

Using the beamformed mixture $\hat{X}_q^{\text{BF}}(t)$ and the estimated target speech in the previous frame (i.e., $\hat{X}_q^{\text{NN}}(t-1)$) as extra input, the DNN-based enhancement model then predicts the target speech at the current time frame:
\begin{align}
    \hat{X}_q^{\text{NN}}(t) = \texttt{RNN}\Big(\mathbf{Y}(t),\hat{X}_q^{\text{BF}}(t),\hat{X}_q^{\text{NN}}(t-1)\Big),
\end{align}
where the RNN can be any causal DNN models that only leverage signal information up to the current processing frame.

We emphasize that, at inference time, the incurred amount of computation due to the inclusion of the proposed auto-regressive connections is very small, as the covariance matrices can be computed in a very efficient, frame-online, cumulative manner and the DNN only has two more input features to model.
From this regard, in our experiments, we consider the same DNN model but without the proposed auto-regressive connections as the main baseline to compare with.

\subsection{Parallel training mechanisms for ARiSE}\label{parallel_training_description}

The auto-regressive connections are essentially a form of recurrent connections in DNNs, and the most straightforward way to train such a model is via back-propagation through time \cite{Courville2016}.
However, we observe that training the model in this way is quite slow, because, during training, the network unfolded across the recurrent connections can be very deep and the gradients need to be back-propagated through the recurrent connections frame by frame.
To deal with this issue,
we propose to modify two existing parallel training approaches \cite{pan24_interspeech, Recurrent-Deep-Stacking} to speed up the training of the auto-regressive model.
Both of them avoid back-propagating gradients through the auto-regressive connections, hence dramatically speeding up the training procedure.

\begin{figure}[t]
    \centering
    \includegraphics[width=0.55\linewidth]{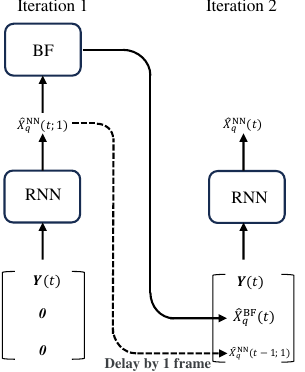}
    \caption{
    Parallel training of ARiSE following PARIS \cite{pan24_interspeech}.
    }\label{parallel_training_figure}
\end{figure}

The first one, which follows the PARIS algorithm \cite{pan24_interspeech} and is illustrated in Figure \ref{parallel_training_figure}, performs feed-forwarding for two iterations at each training step.
In the first iteration, at time frame $t$, the DNN takes in $\big[\mathbf{Y}(t), \mathbf{0}, \mathbf{0}\big]$ as input to obtain an initial target estimate, denoted as $\hat{X}_q^{\text{NN}}(t;1)$, where ``$(\cdot;1)$'' indicates the first iteration.
Based on the initial target estimate, a beamforming result, denoted as $\hat{X}_q^{\text{BF}}$, can be derived based on Equation \ref{MVDR_equation}, and Equation \ref{option_1} or \ref{option_2}.
In the second iteration, the DNN takes in $\big[\mathbf{Y}(t), \hat{X}_q^{\text{BF}}(t), \hat{X}_q^{\text{NN}}(t-1;1)\big]$ as input to obtain the final target estimate.
Following PARIS \cite{pan24_interspeech}, during back-propagation, the DNN in the second iteration is optimized, while the one in the first iteration is not.
That is, the DNN in the first iteration is just utilized to obtain a reasonably accurate $\hat{X}_q^{\text{NN}}(\cdot;1)$ and $\hat{X}_q^{\text{BF}}$ for the second iteration.
At run time, the trained model performs inference in a frame-online auto-regressive manner.
One issue with the PARIS approach is that $\hat{X}_q^{\text{NN}}(\cdot;1)$ and $\hat{X}_q^{\text{BF}}$ are obtained by inputting, at each frame $t$, $\big[\mathbf{Y}(t), \mathbf{0}, \mathbf{0}\big]$ to the DNN in the first iteration, and, as a result, they may not be very accurate, since the two extra input features are both set to zeros and hence would have an input mismatch problem.

\begin{figure}[t]
    \centering 
    \includegraphics[width=1.05\linewidth]{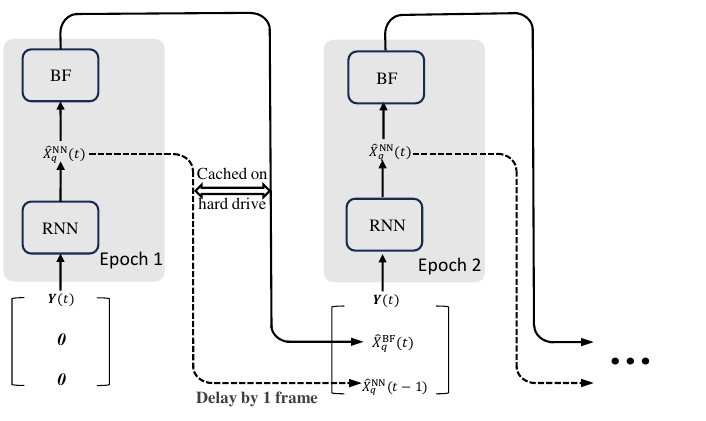}
    \caption{
    Parallel training of ARiSE following RDS \cite{Recurrent-Deep-Stacking}.
    }\label{parallel_training_update_figure}
\end{figure}


The second one, which follows the recurrent deep stacking (RDS) approach \cite{Recurrent-Deep-Stacking} and is illustrated in Figure \ref{parallel_training_update_figure}, caches the predicted target speech and beamforming results of all the training mixtures computed by the DNN model in the previous training epoch, and the cached results are updated after each training epoch.
As the DNN training goes on, the cached intermediate results would become more and more accurate and stable, especially when the model is about to converge.
This way, at inference time, we can safely run the trained DNN in an auto-regressive way for frame-online enhancement.
The RDS approach \cite{Recurrent-Deep-Stacking} can hence alleviate the input mismatch problem in PARIS \cite{pan24_interspeech}.



\begin{table}[t]
\setlength{\tabcolsep}{4pt}
\centering
\caption{IGCRN configuration.}\label{dnn_architecture}
\scalebox{0.9}{
\begin{tabular}{llll}
\midrule
layer name & input size & hyperparameters & output size \\
\midrule 
iGLU($1$)  & $[B,E,161,T]$  & $5\times2, (1,1), 48$  & $[B,48,161,T]$ \\ \midrule
iGLU($2$-$5$)  & $[B,48,161,T]$ & $5\times2, (1,1), 48$ & $[B,48,161,T]$ \\ \midrule
reshape & $[B,48,161,T]$ & - & $[B,161,T,48]$ \\ \midrule
LSTM & $[B \times 161,T,48]$ & $48$ & $[B,161,T,48]$ \\ \midrule
reshape & $[B\times 161,T,48]$ & - & $[B,48,161,T]$ \\ \midrule
iTGLU($5$-$2$) & $[B,96,161,T]$ & $5\times2, (1,1), 48$ & $[B,48,161,T]$ \\ \midrule
iTGLU($1$) & $[B,96,161,T]$ & $5\times2, (1,1), 48$ & $[B,2,161,T]$ \\ \midrule
\end{tabular}
}
\end{table}

\section{Experimental Setup}


We evaluate the proposed algorithms in noisy-reverberant conditions.
The training data and validation data are generated by convolving multi-channel room impuslse responses (RIRs), simulated by using the image method \cite{RIRs}, with clean speech from the Librispeech dataset \cite{librispeech} and noises from the sound effect library\footnote{Available at \url{https://www.sound-ideas.com}.}.
For the test data, the clean speech is from the WSJ0 SI-84 training set \cite{wsj0}, and we randomly draw $150$ utterances from it.
The noises in the test set are from the NOISEX-$92$ dataset \cite{noisex-92}.
We simulate a $6$-microphone uniform circular array with a radius of $8$ cm to generate paired mixture signals and clean target signals.
The array is randomly positioned inside simulated rooms.
There are $4$ noise sources and $1$ target speaker source, totaling $5$ simulated sources.
The reverberation time (T$60$) is randomly sampled from $0.2$ to $1.0$ second.
The signal-to-noise ratio (SNR) is randomly sampled from $-10$ to $10$ dB.
It is also important to note that all the microphones are placed at least $1.0$ m away from the sound source. 
The sampling rate is $16$ kHz.

\begin{table*}
\setlength{\tabcolsep}{3pt}
\centering
\caption{Performance comparison with various benchmarks. Scores are reported in ESTOI/PESQ format.}\label{comparison_benchmarks}
\begin{threeparttable}
\begin{tabular}{c|ccccccccc}
\hline
SNR (dB) & \multicolumn{3}{c|}{-5} & \multicolumn{3}{c|}{0} & \multicolumn{3}{c}{5} \\ \hline
T60 (s) & \multicolumn{1}{c|}{0.3} & \multicolumn{1}{c|}{0.4} & \multicolumn{1}{c|}{0.5} & \multicolumn{1}{c|}{0.3} & \multicolumn{1}{c|}{0.4} & \multicolumn{1}{c|}{0.5} & \multicolumn{1}{c|}{0.3} & \multicolumn{1}{c|}{0.4} & {0.5} \\ \hline
Mixture & 0.513/1.31 & 0.492/1.31 & 0.475/1.28 & 0.640/1.61 & 0.623/1.61 & 0.602/1.57 & 0.767/1.94 & 0.742/1.92 & 0.714/1.88 \\
IGCRN \cite{igcrn} & 0.726/2.05 & 0.704/2.00 & 0.681/1.92 & 0.837/2.53 & 0.822/2.48 & 0.806/2.42 & 0.902/2.92 & 0.886/2.85 & 0.875/2.77 \\
IGCRN-MVDR  & 0.732/2.18 & 0.681/2.11 & 0.644/2.05 & 0.838/2.35 & 0.801/2.27 & 0.767/2.20 & 0.900/2.53 & 0.864/2.43 & 0.837/2.34 \\
ARiSE & 0.805/2.39 & 0.772/2.27 & \textbf{0.747}/2.17 & 0.884/\textbf{2.82} & 0.864/2.71 & 0.844/2.61 & \textbf{0.928/3.13} & \textbf{0.911/3.02} & \textbf{0.899/2.92} \\ 
ARiSE \& finetune & \textbf{0.806/2.41} & \textbf{0.773/2.29} & \textbf{0.747/2.19} & \textbf{0.885/2.82} & \textbf{0.865/2.72} & \textbf{0.845/2.62} & \textbf{0.928/3.13} & \textbf{0.911/3.02} & \textbf{0.899/2.92} \\
ARiSE non-parallel* & 0.738/2.07 & 0.704/1.99 & 0.672/1.90 & 0.846/2.55 & 0.824/2.45 & 0.800/2.39 & 0.908/2.90 & 0.887/2.80 & 0.871/2.69 \\
\hline

\end{tabular}
\begin{tablenotes}
\footnotesize
\item *We only trained the ``ARiSE non-parallel'' system for one week. Therefore, the result does not reflect a fully-converged model. 
\end{tablenotes}
\end{threeparttable}
\end{table*}


The DNN model in ARiSE can be any frame-online enhancement model.
In this paper, we adopt the inplace gated convolution recurrent neural network (IGCRN) \cite{igcrn}, which operates in the T-F domain, as the backbone DNN architecture, considering that it is a light-weight model, which has been shown suitable for real-time, frame-online multi-channel speech enhancement.
IGCRN is an RNN model sandwiched by an encoder-decoder module.
It is configured to use $20$ ms window size and $10$ ms hop size.
Its detailed configuration is provided in Table \ref{dnn_architecture}.
The encoder and decoder respectively consist of $5$ inplace GLUs (iGLU) \cite{igcrn} and inplace Transpose GLUs (iTGLU) \cite{igcrn}.
The shapes of the internal tensors are specified in the format of [\#channels, \#frequencies, \#time\_steps].
The hyperparameter setting is given in the format of (kernel\_size, kernel\_stride, out\_channels) in each convolutional block.
For the LSTM module, we use a unidirectional channel-wise LSTM \cite{igcrn} to ensure causality.

We use the Adam optimizer \cite{AdamAM} for training, starting from an initial learning rate of $0.001$.
The first microphone is designated as the reference microphone.
All the models in this paper are trained to estimate the complex ratio mask \cite{Williamson2016} at the reference microphone by optimizing the $L_1$ loss function, which is defined in the T-F domain on the estimated target speech produced by the DNN.

\begin{table}
\setlength{\tabcolsep}{2.2pt}
\centering
\caption{Ablation results of ARiSE at various T$60$ levels and at $-5$ dB SNR level. Scores are reported in ESTOI/PESQ format.
Best scores are highlighted in bold.}\label{ablation_results}
\fontsize{7.5}{10}\selectfont
\begin{tabular}{c|ccc|ccc}
\hline
 \multirow{2}{*}{ID} & \multirow{2}{*}{\begin{tabular}[c]{@{}c@{}}Parallel\\training\end{tabular}} & \multirow{2}{*}{\begin{tabular}[c]{@{}c@{}}BF\\options\end{tabular}} & \multirow{2}{*}{\begin{tabular}[c]{@{}c@{}}AR\\options\end{tabular}} &  \multicolumn{3}{c}{T60} \\ \cline{5-7}
  &  &  & & \multicolumn{1}{c|}{0.3 s} & \multicolumn{1}{c|}{0.4 s} & {0.5 s}  \\ \hline
 \multicolumn{4}{c|}{Mixture} & {0.513/1.31} & {0.492/1.31} & {0.475/1.28}  \\
\hline
 1 & RDS & Equation \ref{option_2} & $\hat{X}_q^{\text{BF}}$ & 0.791/2.31 & 0.759/2.19 & 0.734/2.10  \\
2 & PARIS & Equation \ref{option_2}  & $\hat{X}_q^{\text{BF}}$ & 0.776/2.26 & 0.744/2.15 & 0.718/2.06  \\
3 & RDS & Equation \ref{option_1}  & $\hat{X}_q^{\text{BF}}$& 0.767/2.19 & 0.737/2.10 & 0.714/2.01  \\
4 & PARIS & Equation \ref{option_1} & $\hat{X}_q^{\text{BF}}$ & 0.746/2.14 & 0.718/2.05 & 0.697/1.97 \\ 
5 & RDS & Equation \ref{option_2} & $\hat{X}_q^{\text{BF}},\hat{X}_q^{\text{NN}}$ & {\textbf{0.805/2.39}} & \textbf{0.772/2.27} & \textbf{0.747/2.17} \\
6 & RDS & Equation \ref{option_2} & $\hat{X}_q^{\text{NN}}$ & 0.739/2.14 & 0.713/2.06 & 0.691/1.99 \\
\hline
\end{tabular}
\end{table}

\section{Evaluation Results}

This section presents several ablation results of ARiSE and compares it with benchmark systems on multi-channel speech enhancement in noisy-reverberant conditions.
We use two popular evaluation metrics, including extended short-time objective intelligibility (ESTOI) \cite{estoi}, which measures speech intelligibility, and perceptual evaluation of speech quality (PESQ) \cite{pesq}, which assesses speech quality.
Evaluation results are reported at various SNR and T$60$ levels.

Table \ref{comparison_benchmarks} compares ARiSE with various benchmark systems.
The \textbf{IGCRN} system is the same as ARiSE but without including any auto-regressive connections.
The \textbf{IGCRN-MVDR} system computes an MVDR-based beamforming result based on the estimated target speech produced by \textbf{IGCRN}.
Both IGCRN and IGCRN-MVDR are popular systems for frame-online multi-channel speech enhancement.
In parallel with our primary work, we also trained a non-parallel system, referred to as \textbf{ARiSE non-parallel}, to provide a baseline for comparison. To ensure fair comparison, we intentionally limit the training duration of ARiSE non-parallel to one week, considering that our proposed \textbf{ARiSE} system requires significantly less training time (under one week) when implemented on an NVIDIA RTX $3090$ GPU.
Furthermore, in the \textbf{ARiSE \& finetune} system, we initialize the model with the optimal parameters from ARiSE and conduct subsequent fine-tuning using non-parallel training.
However, this additional fine-tuning step, this approach produces only slight performance improvement compared to the original ARiSE system. 
From the results, we can also observe that ARiSE produces clearly better performance than IGCRN and IGCRN-MVDR, while only requiring a very small amount of extra computation.

Table \ref{ablation_results} reports the results of ARiSE against the two parallel training methods (described in Section \ref{parallel_training_description}), the two BF options to compute beamforming results (described in Equation \ref{option_1} and \ref{option_2}), and the two auto-regressive connections (i.e., DNN-estimated target speech and beamforming results).
For all the models in Table \ref{ablation_results}, at inference time, they perform enhancement in a frame-online auto-regressive manner. Since our observations revealed that performance differences across systems were not particularly pronounced in high signal-to-noise ratio (SNR) scenarios, we consequently limited our comparison in Table \ref{ablation_results} to the metrics obtained under the SNR = -5 dB condition.
Comparing system $1$ with $2$, and $3$ with $4$, we observe that RDS leads to better performance than PARIS for ARiSE.
Comparing system $1$ with $3$, and $2$ with $4$, we can see that using Equation \ref{option_2} to compute the beamforming results yields better performance than using Equation \ref{option_1}.
Comparing system $5$ with $1$, we observe that using both beamforming results and DNN-estimated target speech as auto-regressive connection produces better enhancement.
Comparing system $6$ with $1$, we can see that using DNN-estimated target speech as auto-regressive connection is not as effective as using beamforming results. This comparison indicates the benefits of including beamforming in multi-channel scenarios in AR models.

\section{Conclusions}

We have proposed ARiSE, a generalized framework that uses auto-regressive connections to improve frame-online multi-channel speech enhancement.
To address the problem that naively training auto-regressive models is too slow, we have designed an effective training method that can train the auto-regressive models in a parallel manner.
Evaluation results in noisy-reverberant conditions show that ARiSE can achieve strong multi-channel speech enhancement performance at a small training cost, compared with benchmark models without the auto-regressive connections.

In closing, we emphasize that the DNN models suitable for real-time speech enhancement need to be light-weight, and this requirement dramatically limits the complexity of the DNN architectures themselves.
The proposed ARiSE algorithm presents a novel solution that can leverage signal processing results (i.e., beamforming) to improve light-weight DNN models, at a cost of slightly increased amount of computation, which is spent in the beamforming module.
The proposed idea of leveraging signal processing techniques to improve light-weight DNNs for real-time speech enhancement, we think, could be a valuable contribution to the research community.


\bibliographystyle{IEEEtran}
\normalem
\vspace{2 em}
\bibliography{mybib}

\end{document}